\pdfoutput=1
\documentclass[aps,reprint,prl,nofootinbib,amsfonts,amssymb,amsmath]{revtex4-2}

\usepackage[T1]{fontenc}
\usepackage[utf8]{inputenc}

\usepackage{lmodern,microtype,mathtools,bm,graphicx,xcolor,physics}
\usepackage[scr=boondoxo,bb=boondox]{mathalfa}

\definecolor{darkblue}{rgb}{0.1,0.1,.7}
\usepackage[colorlinks, linkcolor=darkblue, citecolor=darkblue, urlcolor=darkblue,
linktocpage, breaklinks, bookmarksnumbered,
bookmarksdepth=2]{hyperref}
\usepackage{orcidlink}

\newcommand{\SB}{S_{\mathrm{B}}}
\newcommand{\SE}{S_{\mathrm{E}}}
\newcommand{\SIR}{S_{\mathrm{IR}}}
\newcommand{\PsiUV}{\Psi_{\mathrm{UV}}}
\newcommand{\PsiIR}{\Psi_{\mathrm{IR}}}
\newcommand{\PiT}{\Pi^{\perp}}
\newcommand{\PiL}{\Pi^{\parallel}}
\newcommand{\dhat}{\hat{\delta}}

\def\<#1\>{\expval{#1}}

\renewcommand \.   {\cdot}

\newcommand   \half{\frac 1 2}
\newcommand   \lptl{\raise .8ex\hbox{$^\leftarrow$} \hspace{-9pt} \partial} % derivatives with arrows
\newcommand   \lrptl{\raise .8ex\hbox{$^\leftrightarrow$} \hspace{-9pt} \partial} % derivatives with arrows
\def\be#1\ee{\begin{equation}\begin{aligned}#1\end{aligned}\end{equation}}
\makeatletter
\DeclareRobustCommand\bea{\@ifnextchar[{\@@bea}{\@bea}}
\def\@@bea[#1]#2\eea{\begin{subequations}\begin{align}#2\end{align}\label{#1}\end{subequations}}
\def\@bea#1\eea{\begin{subequations}\begin{align}#1\end{align}\end{subequations}}
\makeatother
\mathcode`l="8000
\begingroup
\makeatletter
\lccode`\~=`\l
\DeclareMathSymbol{\lsb@l}{\mathalpha}{letters}{`l}
\lowercase{\gdef~{\ifnum\the\mathgroup=\m@ne \ell \else \lsb@l \fi}}%
\endgroup
\renewcommand \a  {\alpha}

\newcommand   \g  {\gamma}

\newcommand   \de {\delta}

\newcommand   \e  {\epsilon}
\renewcommand \l  {\lambda}

\newcommand   \SU    {\mathrm{SU}}

\newcommand   \Sym   {\mathrm{Sym}}

\newcommand   \cD {\mathcal{D}}

\newcommand   \cG {\mathcal{G}}

\newcommand   \cK {\mathcal{K}}

\newcommand   \cV {\mathcal{V}}

\begin{document}
\title{Stochastic gauge fixing from the holographic renormalization group}
\author{Soner Albayrak\,\orcidlink{0000-0002-9223-0115}}
\affiliation{Department of Physics, Middle East Technical University, Ankara, Türkiye}

\date{\today}
\begin{abstract}
We extend the dictionary between stochastic quantization and holographic Wilsonian renormalization to Yang–Mills theory in four-dimensional anti-de Sitter space at tree level. The radial component of the bulk gauge field is Zwanziger's gauge-fixing function; radial gauge is Parisi–Wu quantization, whose random walk of gauge modes is the longitudinal bulk propagator. We obtain the triple- and quadruple-trace couplings of gluons at finite cutoff in closed form. On transverse data, their dependence on the bulk gauge resides in a single zero mode, which drops out of the flat-space limit.
\end{abstract}

\maketitle

\subsection{Motivation}

Stochastic quantization \cite{Parisi:1980ys,Damgaard:1987rr} and holography \cite{Maldacena:1997re,Witten:1998qj} both describe a $d$-dimensional Euclidean field theory through a theory with one more dimension. In the former, the fields diffuse in a fictitious time $t$ according to a Langevin equation, and the $d$-dimensional theory governs the equilibrium reached as $t\to\infty$. In the latter, the additional dimension is the radial coordinate $z$ of an anti-de Sitter (AdS) bulk, which geometrizes the energy scale of the boundary theory. This resemblance \cite{Lifschytz:2000bj,Polyakov:2000xc,Mansi:2009mz,Minic:2010pw} was sharpened into a dictionary \cite{Oh:2012bx} between stochastic time evolution and the radial flow of the holographic Wilsonian action \cite{Heemskerk:2010hk,Faulkner:2010jy}: the stochastic time is the radial cutoff, the Fokker--Planck Hamiltonian is the radial Hamiltonian, and the multi-trace couplings of the Wilsonian action are fixed by stochastic correlation functions. This map has been worked out for free scalars, fermions, and Maxwell fields \cite{Oh:2012bx,Jatkar:2013uga,Oh:2013tsa,Oh:2015xva,Moon:2017btx}; and, for interacting scalars \cite{Oh:2021bxx,Lee:2023ynb,Kim:2023bhp,Chae:2024zyn}.

Interacting gauge fields are absent from this list, and even the free analysis \cite{Oh:2012bx} was restricted to transverse modes. This is a conspicuous gap, since non-abelian gauge theories were among the main motivations for stochastic quantization: Parisi and Wu proposed it as a way of quantizing them without fixing the gauge \cite{Parisi:1980ys}, and Zwanziger's stochastic gauge fixing \cite{Zwanziger:1981kg,Baulieu:1981ec} tames the gauge modes without Faddeev--Popov ghosts \cite{Faddeev:1967fc} or Gribov copies \cite{Gribov:1977wm}. What do these constructions correspond to in the bulk?

In this Letter we answer this question for $\SU(N)$ Yang--Mills theory in AdS$_4$ at tree level. First, we recast the all-order tree-level derivation of the dictionary \cite{Oh:2021bxx} as a statement about wavefunctions: the distribution of the Langevin process at time $t$ is the product of the ultraviolet and infrared radial wavefunctions at the cutoff $\e=t$. This identifies the process as the ground-state (Doob) transform of the radial evolution, and in this form the argument extends to gauge fields, whose gauge-fixed drift is not a gradient. Second, the radial component of the bulk gauge field is Zwanziger's gauge-fixing function, so that a choice of bulk gauge is a choice of stochastic gauge fixing (Table~\ref{tab: dictionary}). Third, using the momentum-space technology developed for gluon correlators in AdS \cite{Raju:2011mp,Albayrak:2018tam,Albayrak:2019asr,Albayrak:2026auu}, we obtain the triple- and quadruple-trace couplings of gluons at finite cutoff in closed form, and we show that their entire dependence on the bulk gauge resides in a single longitudinal zero mode. 

In summary, we provide an identification that \textbf{(1)} gives the gauge-fixing function of stochastic quantization a geometric meaning, \textbf{(2)} explains the longitudinal propagator of Witten-diagram computations as a random walk, and \textbf{(3)} carries over to cosmological correlators in temporal gauge. Derivations and further results are given in the companion paper \cite{Albayrak:2026wot}.

\begin{table}[t]
\vspace{-.61em}
\caption{\label{tab: dictionary}Stochastic quantization versus holographic Wilsonian RG at tree level. The first six entries are the dictionary of \cite{Oh:2012bx,Oh:2021bxx}, the sixth being its restatement in terms of wavefunctions; the others are obtained in this Letter.}
\begin{ruledtabular}
\begin{tabular}{ll}
Stochastic quantization & Holographic Wilsonian RG\\
\hline
Fictitious time $t$ & Radial cutoff $\e$\\
Equilibrium action $\SE$ & $2\SIR=-2\log\PsiIR$\\
$\psi_S=P\,e^{\SE/2}$ & $\PsiUV/Z=e^{-\SB}/Z$\\
Hamiltonian $H_{FP}$ & Radial Hamiltonian $H_{RG}$\\
Initial state $P=\de[\Phi]$ & Dirichlet condition at $z=0$\\
Distribution $P[\Phi,t]$ & $\PsiUV\PsiIR/Z$ at $\e=t$\\
\hline
Gauge-fixing function $v$ & Radial component $A_z$\\
Parisi--Wu process & Radial gauge $A_z=0$\\
Zwanziger's $v=\a^{-1}\partial_iA_i$ & Gauge $A_z=\a^{-1}\partial_iA_i$\\
Noise constraint $D_i\eta_i=0$ & Gauss law ($A_z$ integrated)\\
Random walk of gauge modes & Longitudinal propagator
\end{tabular}
\end{ruledtabular}
\end{table}

\subsection{Setup}
Yang--Mills theory is clasically Weyl invariant in four dimensions. In the Poincar\'e patch of Euclidean AdS$_4$, $ds^2=(dz^2+d\bm x^2)/z^2$, its action is therefore the flat-space one, $S=\frac14\int dz\,d^3x\,F^a_{\mu\nu}F^a_{\mu\nu}$ with $F^a_{\mu\nu}=\partial_\mu A^a_\nu-\partial_\nu A^a_\mu+g f^{abc}A^b_\mu A^c_\nu$, on the half-space $z>0$; we impose Dirichlet conditions at the conformal boundary $z=0$. In radial gauge, $A^a_z=0$,
\be
\label{eq: YM radial gauge}
S=\int_0^\infty dz\int d^3x\left[\half\partial_zA^a_i\partial_zA^a_i+\half B^a_iB^a_i\right]\;,
\ee
with $B^a_i=\half\e_{ijk}F^a_{jk}$. The radial ``time'' $z$ does not appear explicitly.

In the holographic Wilsonian renormalization group (RG) \cite{Heemskerk:2010hk,Faulkner:2010jy}, the bulk path integral is split at a cutoff surface $z=\e$,
\be
\label{eq: HP split}
Z=\int\cD A\;\PsiUV[A;\e]\;\PsiIR[A;\e]\;,
\ee
where $\PsiUV\coloneqq e^{-\SB}$ is the path integral over the ultraviolet (UV) region $0<z<\e$, and $\PsiIR\coloneqq e^{-\SIR}$ the one over the infrared (IR) region $z>\e$ with regular behavior in the interior, both at fixed $A(\e)=A$. Here $\SB[A;\e]$ is the Wilsonian boundary action at the cutoff scale $1/\e$. The wavefunctions obey the radial Schr\"odinger equations $\partial_\e\PsiUV=-H_{RG}\PsiUV$ and $\partial_\e\PsiIR=H_{RG}\PsiIR$ \cite{Mansfield:1999kk}, with $H_{RG}=\int d^3x\,[-\half\de^2/\de A^a_i\de A^a_i+\half B^a_iB^a_i]$.

In stochastic quantization, a field $\Phi$ evolves according to the Langevin equation
\be
\label{eq: langevin}
\partial_t\Phi(t,x)=-\half\fdv{\SE[\Phi]}{\Phi(t,x)}+\eta(t,x)\;,
\ee
with Gaussian white noise $\<\eta(t,x)\eta(t',x')\>=\de(t-t')\de^d(x-x')$, and its distribution $P[\Phi,t]$ relaxes to $e^{-\SE}$. The combination {$\psi_S\coloneqq Pe^{\SE/2}$} obeys {$\partial_t\psi_S=-H_{FP}\psi_S$} with the Fokker--Planck Hamiltonian {$H_{FP}=\int d^dx\left[-\half\frac{\de^2}{\de\Phi^2}+\frac18\left(\fdv{\SE}{\Phi}\right)^2-\frac14\frac{\de^2\SE}{\de\Phi^2}\right]$}. The dictionary of \cite{Oh:2012bx,Jatkar:2013uga,Oh:2021bxx} is the map $(t,\SE,\psi_S)\leftrightarrow(\e,2\SIR,\PsiUV)$.\footnote{For Yang--Mills theory, $\SE[A]=2\SIR[A]$ is twice the tree-level generating functional of the
correlators of the boundary current: a non-local, gauge-invariant functional of the
three-dimensional field $A_i$, not a Yang--Mills action.} When written in terms of the kernels of $\SB=\sum_nD^{(n)}\Phi^n$, $\SE=\sum_nG^{(n)}\Phi^n$ and $-\log P=\sum_nP^{(n)}\Phi^n$, it reads
\be
\label{eq: multitrace dictionary}
D^{(n)}(\e)\evaluated_{\e=t}=P^{(n)}(t)-\half G^{(n)}\;.
\ee
We call $D^{(n)}$ multi-trace couplings \cite{Oh:2012bx,Oh:2021bxx}; the couplings of the multi-trace operators in the Wilsonian action of the boundary theory follow from them by a Legendre transformation \cite{Witten:2001ua,Heemskerk:2010hk,Faulkner:2010jy}.

\subsection{Langevin process as a ground-state transform}
Equation \eqref{eq: multitrace dictionary} has been verified one coupling at a time, and derived for all $n$, for scalar theories without explicit radial dependence, by completing the square in the bulk action \cite{Oh:2021bxx}. We recast that argument in a form that applies to gauge fields and makes its probabilistic content explicit. Consider a bulk action $S=\int_0^\infty dz\int d^dx\,[\half(\partial_z\Phi)^2+\cV[\Phi]]$ with $\cV$ independent of $z$, such as \eqref{eq: YM radial gauge}. The shift $z\to z-\e$ shows that the on-shell action $\SIR$ of the region $z>\e$ does not depend on $\e$, so its Hamilton--Jacobi equation reduces to a zero-energy condition \cite{deBoer:1999tgo}. For $\SE\equiv2\SIR$,
\be
\label{eq: stationary HJ}
\frac18\int d^dx\left(\fdv{\SE}{\Phi}\right)^2=\int d^dx\,\cV[\Phi]\;.
\ee
The distribution of the Langevin process \eqref{eq: langevin} started at $\Phi(0)=0$ has the Onsager--Machlup representation \cite{Machlup:1953zz}
\small 
\begin{equation*}
P[\Phi,t]=\int\limits_{\Phi(0)=0}^{\Phi(t)=\Phi}\cD\Phi\exp\bigg(-\half\int_0^tdt'\int d^dx
\Big(\partial_{t'}\Phi+\half\fdv{\SE}{\Phi}\Big)^2\bigg),
\end{equation*}
\normalsize
up to a Jacobian that does not contribute at tree level. The cross term of the square is a total derivative, which integrates to $\half\SE[\Phi]-\half\SE[0]$, and by \eqref{eq: stationary HJ} the remaining terms are the bulk action of the slab $0<z<t$. Hence $Pe^{\SE/2}\propto e^{-\SB[\Phi;t]}$ at tree level, i.e.
\be
\label{eq: P as HP integrand}
P[\Phi,t]=\frac{\PsiUV[\Phi;t]\,\PsiIR[\Phi]}{Z}\;,
\ee
where the normalization of $P$ identifies the constant with the partition function \eqref{eq: HP split}. Equation \eqref{eq: P as HP integrand} implies \eqref{eq: multitrace dictionary} for all $n$. It covers Maxwell and Yang--Mills theory in AdS$_4$ as well as the conformally coupled scalars of \cite{Oh:2021bxx,Lee:2023ynb}.

According to \eqref{eq: P as HP integrand}, the Langevin process samples the bulk field on the slice $z=t$ with the Heemskerk--Polchinski weight. Equivalently, it is the ground-state (Doob) transform \cite{BSMF_1957__85__431_0,Nelson:1966sp,Glimm:1987ng,Dijkgraaf:2009gr,Horava:2008jf} of the radial evolution by $\PsiIR$:  since $\PsiIR$ is annihilated by $H_{RG}$ at tre level, the drift is $\de\log\PsiIR/\de\Phi=-\half\de\SE/\de\Phi$, and the equilibrium weight $e^{-\SE}=\PsiIR^2$ is the square of the regular radial wavefunction. Three consequences follow. The Dirichlet condition at $z=0$ is the sharp initial state $P[\Phi,0]=\de[\Phi]$, and a general UV boundary condition corresponds to the initial distribution $P[\Phi,0]\propto\PsiUV[\Phi;0]\,\PsiIR[\Phi]$, provided it is normalizable. For bulk theories with explicit radial dependence, \eqref{eq: P as HP integrand} with $\PsiIR[\Phi,t]$ still solves a Fokker--Planck equation, with the time-dependent drift $\de\log\PsiIR[\Phi;t]/\de\Phi$ (cf.\ \cite{Chae:2024zyn}). And the relation does not rely on perturbation theory: if $\PsiIR$ is the exact positive ground state of $H_{RG}$, with energy $E_0$, and $\SE\equiv-2\log\PsiIR$, then $H_{FP}=H_{RG}-E_0$ holds identically.

\subsection{Bulk gauge choice as stochastic gauge fixing}
For a gauge field, gauge invariance of $\SE$ implies $D_i\,\de\SE/\de A_i=0$: the drift is orthogonal to the gauge orbits, whereas the noise is not. In the Parisi--Wu process, the gauge modes therefore perform an unrestricted random walk, and $P$ has no normalizable equilibrium along the orbits \cite{Parisi:1980ys}. Zwanziger damped them, without affecting gauge-invariant observables, by adding to the drift a term $D_iv$ tangent to the orbits, e.g.\ with $v=\a^{-1}\partial_iA_i$ \cite{Zwanziger:1981kg}. If $v$ is regarded as the time component $A_t$ of a $(d+1)$-dimensional gauge field, the Langevin equation takes the covariant form \cite{Baulieu:1999wz}:
\be
\label{eq: BZ}
F_{ti}\equiv\partial_tA_i-D_iA_t=-\half\fdv{\SE}{A_i}+\eta_i\;.
\ee
We now show that $A_t$ is the radial component $A_z$ of the bulk gauge field. With $A_z\neq0$ the radial derivative in \eqref{eq: YM radial gauge} is replaced by $F_{zi}$, while the Onsager--Machlup action of \eqref{eq: BZ} is $\half\int\left(F_{ti}+\half{\de \SE}/{\de A_i}\right)^2$. Its cross term is again a total derivative: $\int\partial_tA_i\,\de\SE/\de A_i=d\SE/dt$ as before, whereas $\int(D_iA_t)^a\,\de\SE/\de A^a_i=-\int A^a_t\,(D_i\,\de\SE/\de A_i)^a=0$ by gauge invariance. By \eqref{eq: stationary HJ}, the remaining terms are the bulk Yang--Mills action $\frac14\int F_{\mu\nu}F_{\mu\nu}$ of the slab with $A_z=A_t$. Hence \eqref{eq: P as HP integrand} holds for any $A_t$, even though the gauge-fixed drift is not a gradient. This gives the lower part of Table~\ref{tab: dictionary}: bulk gauge invariance is the $(d+1)$-dimensional gauge invariance of \cite{Baulieu:1999wz}, radial gauge is the Parisi--Wu process, and the ``parabolic'' gauge $A_z=\a^{-1}\partial_iA_i$, with $\a>0$, is Zwanziger's gauge. If $A_z$ is integrated over rather than fixed, its equation of motion, the Gauss law $D_iF_{zi}=0$, formally becomes the constraint $D_i\eta_i=0$ on the noise.

The argument survives in the path integral because the gauge-fixing measure is trivial: in the parabolic gauge, the Faddeev--Popov operator $D_z-\a^{-1}\partial_iD_i$ is first order in $z$ and acts on gauge parameters that vanish at $z=0$, so its determinant is a field-independent constant ($f^{aac}=0$) and no ghosts are needed, as in stochastic gauge fixing \cite{Zwanziger:1981kg,Baulieu:1981ec}. Moreover, the gauge is reached by solving a forward parabolic equation for the gauge parameter, whose solution is unique order by order in $g$: there are no Gribov copies in perturbation theory \cite{Zwanziger:1981kg,Baulieu:1999wz}.

\subsection{Gauge modes as Brownian motion}
We Fourier transform along the boundary, write $k\equiv\abs{\bm k}$, and introduce the projectors $\PiT_{ij}=\de_{ij}-k_ik_j/k^2$ and $\PiL_{ij}=k_ik_j/k^2$. The quadratic kernel of $\SE$ is $G^{(2)}=k\,\PiT\equiv\Omega_{\bm k}$. For a function $W(k)$, we define the dressing operator
\be
\label{eq: K operator}
\cK_{\bm k}[W]\coloneqq\PiT(\bm k)\,W(k)+\PiL(\bm k)\,W(0)=W(\Omega_{\bm k})\;,
\ee
i.e.\ the functional calculus of $\Omega_{\bm k}$, whose eigenvalues are $k$ (transverse) and $0$ (longitudinal); it is an algebra homomorphism, $\cK_{\bm k}[W_1W_2]=\cK_{\bm k}[W_1]\.\cK_{\bm k}[W_2]$. In radial gauge, the free Langevin equation is $\partial_tA=-\Omega_{\bm k}A+\eta$. For $A(0)=0$ its response function is $\cK_{\bm k}[e^{-k\tau}]\,\theta(\tau)$, and its correlator is
\bea[eq: free correlator]
\<A_i(t,\bm k)A_j(t',\bm k')\>&=\dhat(\bm k+\bm k')\,\cK_{\bm k}[\cG_k(t,t')]_{ij}\;,\\
\cG_k(z,w)&=\frac{e^{-k|z-w|}-e^{-k(z+w)}}{2k}\;,
\eea
with $\dhat(\bm k)\coloneqq(2\pi)^3\de^3(\bm k)$. Transverse modes perform Ornstein--Uhlenbeck processes, while longitudinal modes perform Brownian motion with covariance $\cG_0(t,t')=\min(t,t')$. Equation \eqref{eq: free correlator} is the bulk-to-bulk propagator of gluons in radial gauge \cite{Raju:2011mp,Albayrak:2018tam}: its longitudinal part, familiar from Witten-diagram computations \cite{Albayrak:2019asr}, is the random walk of the gauge modes in Parisi--Wu quantization. The same structure appears in the double-trace couplings. The UV region gives $2 D^{(2)}(\e)=\cK_{\bm k}[k\coth k\e]=\PiT k\coth k\e+\PiL/\e$ while the Langevin process gives $P^{(2)}(t)=\half\<AA\>^{-1}=\cK_{\bm k}[k/(1-e^{-2kt})]$, so that \eqref{eq: multitrace dictionary} holds in both sectors. The longitudinal coupling $1/(2\e)$, found for Maxwell fields in \cite{Heemskerk:2010hk,Faulkner:2010jy}, is half the inverse variance of a Brownian path. In Zwanziger's gauge, the gauge modes become Ornstein--Uhlenbeck processes with rate $\gamma_k\coloneqq k^2/\alpha$, and the longitudinal double-trace coupling becomes $\gamma_k/\left(1-e^{-2\g_k\e}\right)$, which flows exponentially fast from $1/(2\e)$ in the UV to the gauge-fixing term $k^2/\a$ in the IR.

\subsection{Multi-trace couplings at finite cutoff}
In radial gauge the cubic and quartic terms of \eqref{eq: YM radial gauge} contain only boundary derivatives, so the vertices $V$ and $V^{(4)}$ are $z$-independent tensors and all radial dependence sits in the propagators \cite{Raju:2011mp}. Since $\cK$ is a homomorphism, every tree diagram, on either side of the dictionary, equals the vertex tensors contracted with $\cK$-dressed radial integrals of a scalar theory with the same topology \cite{Albayrak:2019asr}. In the UV region these are built from the bulk-to-cutoff propagator $S_k(z)$ and bulk-to-bulk propagator $\cG^{\rm UV}_k(z,w)$ of the slab, and schematically read
\bea[eq: D3 D4]
\hspace{-.78em}
D^{(3)}(\e)&=\frac{1}{3!}\,V\circ\cK_1\cK_2\cK_3[I_3(\e)]\;,\\
\hspace{-.78em}
D^{(4)}(\e)&=\frac{1}{4!}V^{(4)}\circ\cK[I_4(\e)]-\frac18\Sym\big[VV\circ\cK[F(\e)]\big]
\eea
with $\cK_r\coloneqq\cK_{\bm k_r}$ dressing the external leg $r$, the symbol $\circ$ contracting the vertex indices with the dressing operators, and $\Sym$ averaging over permutations of the external legs.\footnote{For more details on the notation and explicit analytic expressions, we refer the reader to the companion paper \cite{Albayrak:2026wot}.} The scalar seeds are $I_n(\e)=\int_0^\e dz\prod_{r=1}^nS_{k_r}(z)$ and $\displaystyle F(\e)=\!\int_0^\e\!\!\!\int_0^\e\!\!dz\,dw\,S_{k_1}(z)S_{k_2}(z)
\cG^{\rm UV}_{s}(z,w)S_{k_3}(w)S_{k_4}(w)$. The contact seeds are those of conformally coupled scalars \cite{Oh:2021bxx,Lee:2023ynb}, whereas the exchange seed $F$ and its longitudinal limit $F^{(0)}\equiv F|_{s\to0}$ are new.

On the stochastic side, $P^{(3)}$ and $P^{(4)}$ follow from the connected three- and four-point functions of the Langevin process, which are built from the response function, the correlator \eqref{eq: free correlator} and the vertices of $\SE$. We find that they are given by \eqref{eq: D3 D4} with the seeds shifted by their infrared values, $I_n(t)\to I_n(t)+1/k_{1\cdots n}$ and $F(t)\to F(t)+1/[k_{1234}(k_{12}+s)(k_{34}+s)]$, where $k_{1\cdots n}\equiv k_1+\dots+k_n$. This is \eqref{eq: multitrace dictionary}, since the kernels of $\SE=2\SIR$ are twice the $\e\to\infty$ limit of \eqref{eq: D3 D4}. All couplings with $n\geq3$ vanish as $\e\to0$, consistent with the sharp initial state. 

For transverse polarizations, $\bm\e_r\.\bm k_r=0$ with \mbox{$e_{rs}\coloneqq \bm \e_r\. \bm \e_s$}, the fourth variation of $\SB$ decomposes as $\mathscr{D}_4(1,2,3,4;\e)\,f^{a_1a_2c}f^{ca_3a_4}+\mathscr{D}_4(1,3,2,4;\e)\,f^{a_1a_3c}f^{ca_2a_4}$ \cite{DelDuca:1999rs}, with the colour-ordered flows
\bea[eq: colour ordered flows]
\hspace{-.6em}\mathscr{D}_4(1,2,3,4;\e)&=g^2\big[N_s(\e)-N_t(\e)+C_{1234}\,I_4(\e)\big]\;,\\
\hspace{-.6em}N_s(\e)&=n_s^\perp F_s(\e)+n_s^\parallel F^{(0)}_s(\e)\;,
\eea
where $F_s$ is the exchange seed for pairing $(12)(34)$, $N_t$ follows from $N_s$ by cyclic relabelling $2\to4\to3\to2$, and
\begin{align*}
n_s^\perp&=v_{12}\.\PiT(\bm k_1+\bm k_2)\.v_{34}\;,\\
n_s^\parallel&=-e_{12}e_{34}(k_1^2-k_2^2)(k_3^2-k_4^2)/s^2\;,\\
v_{12}&=e_{12}(\bm k_1-\bm k_2)+2(\bm\e_1\.\bm k_2)\,\bm\e_2-2(\bm\e_2\.\bm k_1)\,\bm\e_1\;,\\
C_{1234}&=2e_{13}e_{24}-e_{12}e_{34}-e_{14}e_{23}\;.
\end{align*}
As $\e\to\infty$, \eqref{eq: colour ordered flows} reduces to the colour-ordered gluon correlators of \cite{Albayrak:2019asr}, since $\SIR$ generates the tree-level correlators. Its residue at the total-energy pole $k_{1234}\to0$ must be the flat-space colour-ordered amplitude \cite{Maldacena:2011nz,Raju:2012zr}: we have checked numerically that it vanishes for the all-plus and single-minus helicity configurations and reproduces the Parke--Taylor formula \cite{Parke:1986gb} for the MHV ones.

\subsection{Gauge dependence and the Gauss law}
On transverse data, the triple-trace couplings are the same in Zwanziger's gauge for every $\a$. The quadruple-trace couplings are not, because the internal line of the exchange diagrams can be longitudinal. This channel is special already in radial gauge: all seeds approach their infrared values exponentially fast, except for the longitudinal one,
\be
\label{eq: F0 power law}
F^{(0)}(\e)=\frac{1}{k_{1234}k_{12}k_{34}}-\frac{1}{\e\,k_{12}^2k_{34}^2}+\dots\;,
\ee
where the dots are exponentially small. To understand this power law, let $W_{12}\equiv\partial_zS_{k_1}\,S_{k_2}-S_{k_1}\,\partial_zS_{k_2}$. Up to normalization, $e_{12}W_{12}$ is the colour-charge density $gf^{abc}A^b_i\partial_zA^c_i$ of two transverse modes, i.e.\ the source of the Gauss law. Since $\partial_zW_{12}=(k_1^2-k_2^2)S_{k_1}S_{k_2}$, an integration by parts against the longitudinal propagator $\cG^{\rm UV}_0=z_<(\e-z_>)/\e$, the covariance of a Brownian bridge, gives
\small
\begin{equation*}
n_s^\parallel F_s^{(0)}={}-\frac{e_{12}e_{34}}{s^2}\bigg[\int_0^\e\!dz\,W_{12}W_{34}
-\frac1\e\int_0^\e\!dz\,W_{12}\int_0^\e\!dw\,W_{34}\bigg].
\end{equation*}
\normalsize
The first term is the Coulomb energy of the radial colour charges. It is local in $z$, i.e.\ instantaneous in stochastic time, and it alone carries the total-energy pole in the infrared. The second term is the zero mode of the bridge: in radial gauge, the internal longitudinal field vanishes at both ends of the slab, so that $E^\parallel=\partial_zA^\parallel$ has zero mean and cannot follow the total charge. The power law \eqref{eq: F0 power law} comes entirely from this term; it is the unrestricted random walk of the gauge modes of the Parisi--Wu process.

In Zwanziger's gauge the Coulomb term is unchanged and only the zero mode changes \cite{Albayrak:2026wot}. With $\gamma_s\coloneqq s^2/\alpha$, it now relaxes exponentially, but to an $\a$-dependent value,
\bea[eq: D4 alpha IR]
\hspace{-0.8em}\mathscr{D}^{(\a)}_4(1,2,3,4;\infty)=&\mathscr{D}_4(1,2,3,4;\infty)+g^2\big(Z_s-Z_t\big),\\
\hspace{-.8em}
Z_s=&{}\frac{2e_{12}e_{34}}{\a}\,\frac{k_1-k_2}{k_{12}+\g_s}\,\frac{k_3-k_4}{k_{34}+\g_s}\;,
\eea
with $Z_t$ obtained by the same relabelling. Since $Z_s$ has no total-energy pole, the flat-space limit does not depend on $\a$. By \eqref{eq: multitrace dictionary} the same shift appears in the equilibrium weights: beyond quadratic order, the equilibrium distribution of Zwanziger's process is not the naive gauge-fixed weight $e^{-\SE-\a^{-1}\int(\partial_iA_i)^2}$, as expected for a gauge-fixing force that is not conservative \cite{Baulieu:1999wz}, and \eqref{eq: D4 alpha IR} is the first correction. We have also derived this shift on the stochastic side alone, from the stationary Fokker--Planck equation of Zwanziger's process, which checks the gauge dictionary at quartic order for a drift that is not a gradient.

Finally, the zero mode is absent when the Gauss law is imposed: it vanishes as $\a\to0$ when the gauge condition becomes the Coulomb gauge $\partial_iA_i=0$ with $A_z$ fixed by the Gauss law, and also if $A_z$ is integrated over. In fact, minimizing the Wilsonian action of any bulk gauge over gauge rotations $U$ of the cutoff data gives the Gauss-law action, $S^{\rm G}_{\rm B}[A]=\min_US_{\rm B}[A^U]$, so the zero mode is a redundant coupling \cite{Nickel:2010pr}. The redefinition that removes it is non-local; however, it is the radial-gauge kernels that have the locality expected of a Wilsonian action, as for Maxwell fields \cite{Heemskerk:2010hk,Faulkner:2010jy}. The gauge-independentg part of the quadruple-trace flows is thus the transverse exchange, the contact term, and the Coulomb term.

\subsection{Closing Remarks}
At tree level, the holographic Wilsonian flow of gluons in AdS$_4$ is a Langevin process in the radial direction, and the freedom of choosing the bulk gauge is the freedom of stochastic gauge fixing. Beyond tree level, the identity $H_{FP}=H_{RG}-E_0$ reduces the dictionary to the definition of the exact $\PsiIR$; if $\SE$ is kept at its tree-level value instead, $H_{FP}$ and $H_{RG}$ differ by the one-loop potential $-\frac{1}{4}\int\de^2\SE/\de\Phi^2$ \cite{Albayrak:2026wot}. The trace anomaly then makes $\PsiIR$, and hence the drift, depend on the cutoff, as is the case already at tree level for gravitons  \cite{Albayrak:2019yve,Albayrak:2020bso}.

Finally, under the continuation to de Sitter space \cite{Maldacena:2002vr}, $\PsiIR$ becomes the Bunch--Davies wavefunction and $e^{-\SE}=\PsiIR^2$ the Born-rule weight $\abs{\Psi_{\rm BD}}^2$, so the equilibrium correlators of the Langevin process are the Euclidean counterparts of in-in correlators. Temporal gauge, the standard choice for cosmological gluon correlators \cite{Albayrak:2026auu}, is then Parisi--Wu quantization, and a stochastic treatment of gauge potentials during inflation, in the manner of stochastic inflation for light scalars \cite{Starobinsky:1986fx,Starobinsky:1994bd,Gorbenko:2019rza}, would have to deal with the same zero mode, Gauss law and damping gauges \cite{Albayrak:wip}.

\subsection{Acknowledgments}
A first draft of this Letter was condensed from the companion paper \cite{Albayrak:2026wot} with Claude Fable 5.2 (Anthropic), at the author's direction; the author has checked and edited all text and formulas and takes full responsibility for the content. In the early stages of this research, the author was supported by a VIDI grant of the Netherlands Organisation for Scientific Research (NWO) that is funded by the Dutch Ministry of Education, Culture and Science (OCW). In the later stages, he was supported by the BAGEP Award of the Science Academy in Türkiye and by the TUBITAK (The Scientific and Technological Research Council of
Türkiye)\footnote{Please see ROR ID: \href{https://ror.org/04w9kkr77}{https://ror.org/04w9kkr77}} 2232-b International Fellowship for Early Stage Researchers program, project number 122c153.

\subsection{Data availability}
The Wolfram Language file that derives and checks the formulas of this Letter and of the companion paper is openly available \cite{Albayrak:2026wot}.

\bibliographystyle{utphysModified}
\bibliography{references}
\end{document}